# PERFORMANCE ANALYSIS OF CONTENTION WINDOW CHEATING MISBEHAVIORS IN MOBILE AD HOC NETWORKS

R. Kalaiarasi[1], Getsy S. Sara[2] S. Neelavathy Pari[3] and D. Sridharan[3]

[1, 2, 4] Department of Electronics and Communication Engineering, CEG Campus
[3] Department of Information Technology, MIT Campus
Anna University Chennai, India.
[1]*kalaiarasidr@gmail.com*, [2] *getsysudhir@gmail.com*, [3]*neela_pari@yahoo.com*
[4]*Sridhar@annauniv.edu*,

## ABSTRACT

*Mobile Ad Hoc Network (MANET) is a collection of nodes that can be rapidly deployed as a multi-hop network without the aid of any centralized administration. Misbehavior is challenged by bandwidth and energy efficient medium access control and fair share of throughput. Node misbehavior plays an important role in MANET. In this survey, few of the contention window misbehavior is reviewed and compared. The contention window cheating either minimizes the active communication of the network or reduces bandwidth utilization of a particular node. The classification presented is in no case unique but summarizes the chief characteristics of many published proposals for contention window cheating. After getting insight into the different contention window misbehavior, few of the enhancements that can be done to improve the existing contention window are suggested. The purpose of this paper is to facilitate the research efforts in combining the existing solutions to offer more efficient methods to reduce contention window cheating mechanisms.*

## 1. INTRODUCTION

Mobile Ad Hoc Networks (MANET) [22] has recently been the topic of extensive research. The interest in such network stems from their ability to provide temporary and instant wireless networking solutions. MANET consumes huge amount of power and bandwidth and undergoes frequent topology changes [20], [21]. MANET is a self-organizing multi-hop wireless network with no predefined infrastructure. The self-organizing features of rapid deployment make MANET very attractive in military and earth quakes where fixed infrastructure no available [22]. In mobile Ad hoc networks all the nodes have to cooperate to ensure successful communication. An individual node stops cooperating in order to increase its gains such as higher throughput or extended battery life. Such nodes are called selfish or misbehaving nodes. A selfish node may obtain an unfair throughput share by not obeying the (Medium Access Control) MAC mechanism. Such benefits can be obtained by selecting small backoff, not doubling the Contention Window (CW), changing default interframe times and manipulating Network Allocation Vector (NAV) etc. Protocol misbehaviors have been studied in various scenarios in different communication layers and under several mathematical frameworks. Most notably, a heuristic set of conditions is proposed in [1] for testing the extent to which MAC protocol parameters have been manipulated.

A selfish node [2], [3] affects the network operation specifically by the proper selection of backoff intervals but refuse to forward data packets. Denial of Service (DoS) attack can be launched against any layer in the network protocol stack. Malicious nodes can advertise incorrect routing updates in the network or drop all the packets passing through them. This type of attack is called Denial of Service Attack. Selfish node in the MAC layer will try to maximize its own throughput and keep the channel busy. As a side effect of this behavior, legitimate nodes cannot use the channel for transmission.

An attacker may exploit this feature by asserting large duration field, thereby preventing well behaved clients from gaining access to the channel. This type of attack is called Network Allocation Vector (NAV) attack. A malicious node aims primarily to disrupt the normal operation of the network. These nodes continuously send data to each other in order to deplete the channel capacity in their vicinity and hence prevent other legitimate users from communicating [4]. A host exploits this vulnerability and completely cooperates in forwarding data packets but maliciously forces the forwarding operation to fail. This attack mainly targets the route discovery process in order to route the packets through longer routes. The attack also targets flows that traverse through a malicious node, thus forcing the routing protocol to reroute packets around the misbehaving node [5].

## 2. RELATED WORKS

Sundaramurthy and Royer suggest the Asymmetric key Distribution-Mixnets (AD-MIX) [6] protocol to discourage selfishness by nodes in terms of forwarding data packets. Concealing the true destination of a packet from intermediate nodes encourages data forwarding. This forces a node to participate or risk dropping packets that may be destined for the node itself [17].

Incentive compatible medium access control ICMAC [7] is a Time Division Multiple Access (TDMA) based MAC protocol that is resilent to contention window misbehavior. Through a game theoretic approach and the use of the Vickrey auction mechanism, the authors provide incentives for the nodes to cooperate. The TDMA nature of this protocol makes it more complicated to use in ad-hoc environments [12].

In minimax detection [8] framework is employed to analyze the instance of theoretical worst-case attacks. This approach is more robust but no operational method is used to detect misbehavior. Only the successful transmissions are observable, and in the event of collision, it is not possible to determine what terminals were involved in it [13].

A new type of vulnerability was presented by Guang and Assi [9] where a host maliciously modify the protocol timeout mechanism by changing SIFS (Short Inter Frame Space) parameter in 802.11 and cause MAC frames to be dropped at well-behaved nodes [11].

M. Raya, J. P. Hubaux, and I. Aad [10] presented a detection system called DOMINO that does not require any modifications to the MAC protocol and they presented several procedures for detecting misbehaviors that aim at altering protocol parameters such as shorter than Distributed Inter frame Space (DIFS), oversized NAV, and backoff manipulation. The system is implemented at Access Point (AP) and the AP is assumed to be trusted. Traffic traces of sending hosts are collected periodically during short intervals of time called monitoring periods. Collected data is then passed to the DOMINO algorithm. When a node misbehaves its corresponding cheat counter exceeds a certain threshold [14].

## 3. Techniques Proposed For Detecting Selfishness in MANET

### 3.1. DREAM - Detection and REAction Timeout MAC Layer Misbehavior

Lei Guang et al [11] proposed a new type of malicious behavior TimeOut (TO) attack and provided the corresponding detection and reaction schemes. Rather than just correctly identifying the misbehaved nodes, they developed a two stage reaction mechanism when the first reaction stage is to mitigate and second reaction stage is to punish It improves the network performance in the presence of well behaved nodes. Detection schemes should be implemented for a receiver and a transmitter. DREAM system considers being a network that can be either working in an ad hoc or infrastructure mode. Figure 1 shows the detection and first reaction of DREAM method. It considers two types of misbehaved node.

1) misbehaved node that has no knowledge of the DREAM.
2) misbehaved node that has full knowledge of the DREAM.

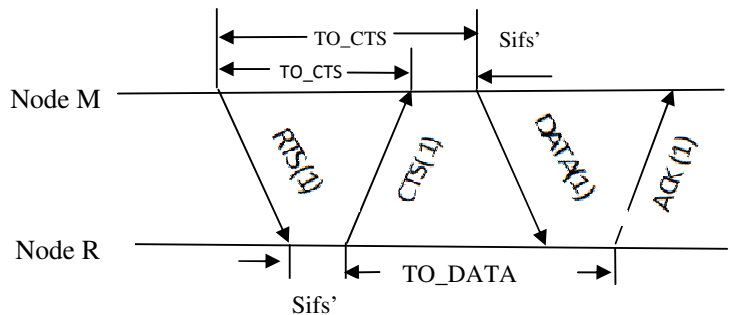

Figure1. Detection and first reaction

Misbehaving nodes are carried through one TO attack. Detection function can be carried out in three steps for misbehaved transmitter and misbehaved receiver.

Receiver node R may receive subsequent Request To Send (RTS) during the data timeout, default timeout of the DATA frame ($TO_{DATA}$ ) or after the timeout. Node R also maintains a parameter called badCredit to evaluate the trustworthiness of every node. When the second RTS arrives at node R during $TO_{DATA}$, it is more likely that node M is misbehaving, R will punish node M by increasing its badCredit parameter heavily.

The adjustment scheme is used to ensure correct misbehavior diagnosis by node R. Once node M is designated as a suspect, then node R will react by expediting the transmission of the Clear To Send (CTS) frame. Node M is not reliable for any future communication. At this point, the trust level of node M is reduced by node R. This monitoring and reacting process continues for a pre-set monitoring period until the trust level of node M falls below a trust level threshold and node R invokes its second reaction scheme.

A node will be marked as a suspect when its badCredit is above a predetermined credit Threshold during a short term monitoring period. Once node is identified as s Suspect Node (SN), the system needs to determine whether the SN is an Untrused (UN) or Trusted Node (TN). For this reason, a long term monitoring is triggered during which the trust level of every SN will be evaluated based on its cooperation. Every time the SN misbehaves, its trust level is reduced and a trust threshold is defined below which the SN is considered as an UN. Based on the different value of trust level, a well-behaved node (WN) can invoke corresponding punishment methods. Correct detection ratio is 100% and misdetection ratio is 20%. In the first reaction scheme there is a slight decrement compared with normal case. In the second reaction scheme network delivery ratio increases. Average delay is less compared with the normal case.

### 3.2. Enhanced Distributed Channel Access (EDCA)

Szymon et al [12] proposed the impact of contention window misbehavior on single hop ad-hoc networks. Uplink and down link traffics were simulated. Throughput, delay, fairness was considered for Transmission Control Protocol (TCP) and User Datagram Protocol (UDP). In this paper author proposed a new distributed channel access mechanism called EDCA. It separates traffic into four access categories (AC) of different priority. Each category has its own set of parameters. Arbitration Inter Frame Space (AIFS), Transmission Opportunity (TXOP), $CW_{min}$, $CW_{max}$. These parameters are responsible for traffic differentiation.

Figure 2 shows the channel access prioritization of EDCA method similar to 802.11 DCF. AIFS [AC] is a parameter which replaces the DIFS of Distributed Coordination Function (DCF). An internal virtual collision mechanism is used to determine which frame can be sent. The potential benefits of misbehaving nodes are measured for UDP/TCP traffic in both uplink and downlink directions.

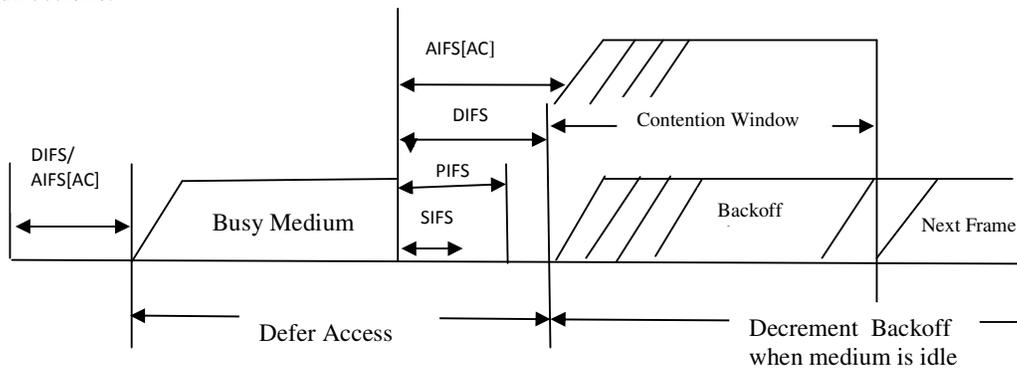

Figure 2. Channel access prioritization

In the uplink scenario, the good nodes had unaltered contention window parameters: $CW_{min}=31$, $CW_{max}=1023$. The bad node had these parameters significantly decreased: $CW_{min}=1$, $CW_{max}=5$. In the downlink scenario, the bad node sends TCP-ACK packets. Therefore, two TCP flows, with an offered load of 8Mbits/s each, were used to put the network in a state of saturation. The bad node increases its throughput until saturation is achieved.

The main conclusion is that CW misbehavior leads to severe unfairness in the uplink direction. The misbehaving node can dominate uplink traffic in terms of both throughput and delay. Misbehaving nodes increasing its throughput is higher for UDP than TCP and more significant for smaller network sizes. A disadvantage of this method is that it gives higher throughput only in the smaller network sizes.

### 3.3. Kolmogorov-Smirov (K-S) Test

Alberto Lopez Toledo and Xiaodong Wang [13] developed nonparametric batch and sequential detectors based on the Kolmogorov-Smirnov (K-S) statistics that do not require any modification on the existing Carrier Sense Multiple Access with Collision Avoidance (CSMA/CA) protocols. The performance of proposed detectors was compared with the optimum detectors with perfect information about the misbehavior strategy for both the batch case and sequential case. Detection can be carried out in three steps:

### 3.3.1. Hypothesis

To detect a misbehaving node observing the sequence of operation in the network. Authors have proposed two hypotheses $H_0$ and $H_1$. $H_0$ corresponds to the observed terminal not misbehaving. $H_1$ corresponds to the terminal misbehaving. $f_0$ and $f_1$ are the probability distributions of the observations when a node is not misbehaving and misbehaving respectively. If the misbehavior is detected, the observer terminals have a mechanism to inform the rest of the network. The sequence of backoff intervals is selected by a terminal. Successful transmissions can be calculated using the number of idle slots between two consecutive transmissions. The number of idle slots can be calculated as $X_i = t_i - t_{i-1} - T_{DIFS} - T_o/\sigma$, $i>1$, where $T_{DIFS}$ is the duration of the DIFS frame, $\sigma$ is the duration of an idle slot, and $T_o$ is the duration of transmissions from other terminals and collisions, including their interframe time

### 3.3.2. Probability Distribution of Legitimate Terminals

Let $p_c$ be the probability of collision, U denotes the uniform probability distribution. Transmission will be successful with probability $(1 - p_c)$. If there is a collision, with probability $p_c$, then the terminal would double its window size and make another attempt after $T_2 \sim U[0, 64]$ slots. If the last transmission is successful then the number of idle slots after the last successful transmission is $xi = T_1 + T_2 \sim U[0, 32] + U[0, 64]$ with probability $p_c(1 - p_c)$. From the above argument one can easily obtain the distribution of the number of idle slots between successful transmissions. In order to characterize and quantify misbehavior, compare $f_1$ to the strategy of a saturating legitimate node $f_0$. . If the terminal is not saturating, its cumulative distribution functions (cdf) satisfies $F_1(x) < F_0(x)$, $\forall x$. If the cdf of a terminal is always on or below the cdf of a well behaved terminal that is always transmitting, then the terminal is definitely not misbehaving.

To obtain Collision Probability Estimation, the distribution $f_0$ of the idle slots between successful transmissions for saturating legitimate terminal and the probability of collision in the network $p_c$ has to be estimated. A terminal can keep track of its own transmissions and count how many of them resulted in collisions. The K-S test determines whether the underlying distribution $f_1$ differs from a hypothesized distribution $f_0$. The K-S test compares the cdf $F1$ obtained from the data samples with the hypothesized cdf $F_0$, and determines whether $F_1 = F_0$, or $F_1 < F_0$, or $F_1 > F_0$. For the misbehavior detection problem, define the null hypothesis as the event where a node is not misbehaving. Choose $H_0 : F_1 \leq F_0$ (not misbehaving), $H_1 : F_1 > F_0$ (misbehaving).

### 3.3.3. N-trunked sequential test

Using the N- trunked sequential test fix the desired false alarm probability of the sequential test to $P_{FA} = \alpha$. Because the sequential test is composed of $N$ tests, we need to calculate the false alarm probability of each stage in order to meet the overall $P_{FA}$.

The K-S detector is able to detect the misbehaviors very fast, requiring less than twice the samples needed by the optimum detector with perfect information. Performance of the K-S detector starts to degrade if $CW_{min} > 29$. It is the main drawback of this method.

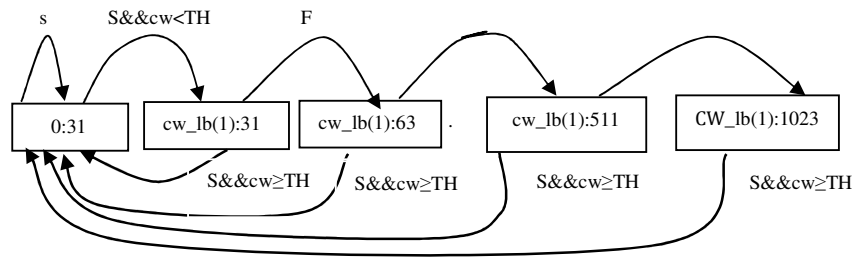

Figure 3. Predictable Random Backoff

### 3.4. PREDICTABLE RANDOM BACKOFF (PRB)

Lei Guang and Chadi Assi [14] proposed Predictable Random Backoff (PRB) algorithm based on modifications of IEEE 802.11 Binary Exponential Backoff (BEB) and forces each node to generate predictable random backoff intervals. Based on PRB, selfish node applies two consequences to manipulate the selection of CW. If the selfish node follows PRB, the negative impact will be mitigated regardless of attack strategies. If the selfish node does not follow PRB, the backoff selection is predictable; the receiver can easily detect the misbehavior of the receiver with direct evidence and perform immediate punishment. Figure 3 shows that the CW continues to increase until it reaches $CW_{MAX}$.

Receiver is used to detect a selfish misbehavior. A Rx can compute actual backoff ($B^i_{act}$) for each received frame. If $B^i_{act} < B^i_{lb}$ labelled as misbehaved Tx where $B^i_{lb}$ is the lower bound backoff computed based on $CW^i_{lb}$. The value of $CW^i_{lb}$ is deterministic and can be easily calculated by the Rx through monitoring the transmissions of Tx. If the selfish node selects larger CW the detection can be mitigated by using $B^i_{act} > B^i_{ub}$. The Author compared the performance of BEB and PRB based on no attack and attack case. The fairness index of PRB ensures a much better fair share of the channel bandwidth when the traffic load becomes higher. In PRB the throughput of the selfish flow has decreased 74%. Each well behaved flow has increased nearly 170%. Compared to BEB, PRB gives higher throughput. PRB achieves better performance than BEB especially in a congested environment. The main drawback of this method is that the attack requires manipulating CW only once and it can intentially choose CW between $B_{lb}$ and $B_{ub}$.

### 3.5. EIED Backoff Algorithm

Nah-Oak Song et al [15] proposed algorithm called Exponential Increase Exponential Decrease (EIED) which is significantly improving the network performance over BEB. In Multiple Increase Linear Decrease (MILD), the contention window size is multiplied by 1.5 on a collision but decreased by 1 on a successful transmission. MILD performs well when the network load is steadily heavy. It does not perform well when the network load is light. In EIED, whenever a packet transmitted from a node is involved in a collision, the contention window size for the node is increased by backoff factor $r_I$, and the contention window for the node is decreased by backoff factor $r_D$ if the node transmits a packet successfully. $r_I = r_D = 2$ was presented in [16]. The EIED backoff algorithm can be represented as follows.

CW=min [$r_I$.CW, CWmax] on a collision,
CW=max [CW/$r_D$, CWmin] on a success.
BEB and MILD are compared with four different cases of EIED. The four cases of EIED always give higher throughput than BEB. The delay performance of MILD is poor when the number of nodes is small. EIED works well specially in a congested environment.

## 3.5. Deterministic and Statistical Method

Venkata Nishanth Lolla et al [17] proposed a combination of deterministic and statistical methods that allow nodes to discern violations of backoff timers by neighboring nodes. First the nodes are exchange the state of their pseudo random number generators with their neighbors. In certain scenarios, a node may not be able to accurately monitor the backoff countdown of a neighbor. In order to compute the expected backoff time, use a statistical online estimation. The tagged node announces the state of its pseudo-random sequence generator using which the monitoring neighbor can determine the sequence of backoff times to be used by tagged node.

The monitoring neighbor may not be able to deterministically determine if the tagged node is using a legitimate backoff countdown process. In such cases, the observed backoff times would differ from what the monitoring node computes using the announced pseudo random sequence generator state. The monitoring neighbor then uses a hypothesis test based on its online estimates of the probabilities $P_{I/B}$ and $P_{B/I}$ to determine if the difference between the observed and expected backoff times is sufficient to deem the tagged node as a misbehaving node.

On line estimation of the system state described by the traffic intensity (p) experienced by a monitoring node is computed based on the number of busy/idle slots observed on the terminal. All nodes use a sequence of backoff timers generated by a pseudo-random number generator (PRNG). The nodes are required to provide their respective MAC addresses as the seed to the PRNG.

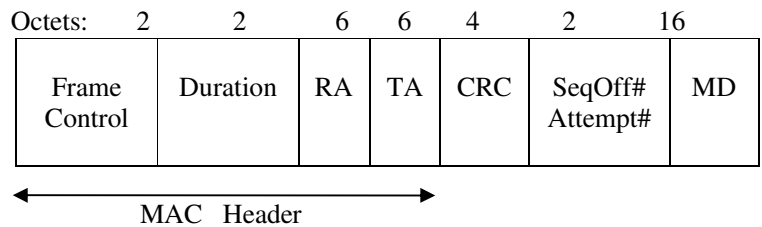

Figure 4. Modified packet structure for RTS.

As shown in Figure 4, every RTS packet sent by the sender S will have the sequence offset number (seqoff#) and an attempt number (Attempt#) included in a new field introduced in the packet. The sender uses 'attempt number' for handling packet retransmissions. It is set to one after every successful transmission by the sender and is incremented by one after every unsuccessful attempt. The sender does not cheat on attempt number .A Message Digest (MD) of the corresponding DATA packet is computed using a hash functions MD5 which is attached as a new field in the RTS packet. During retransmissions, if the receiver notices that a MD for a particular DATA packet matches for multiple retransmissions and the attempt number does not increase with successively received transmissions, the sender is deemed to be misbehaving.

This scheme detects misbehaves by reducing its computed backoff to approximately 75% of the dictated time. The accuracy improves considerably and the misdiagnosis probability reduces. The advantage of this approach is viable even if the load in the network were to be varied.

### 3.6. Hop-by-hop Efficient Authentication Protocol (HEAP)

Rehan Akbani et al [18] proposed a hop-by-hop efficient authentication protocol called HEAP. HEAP authenticates packets at every hop by using modified MAC-based algorithm along with two keys. Every node shares a pair-wise secret hash key, called okey, with each of its neighbors and generates one common secret hash key called okey and securely distributes it to its entire one hop neighbors. A node generates a new Hash Functions for Message Authentication (HMAC) for every individual neighbor using its okey.

HMAC is computed as HMAC (M,K) = H(K x opad|H (K × ipad|M)). Packet index numbers are included in the packet to protect against message replays. HEAP latency is very low compared to TELSA, LHAP and Lu. It provides significant effect on throughput and delivery ratio. Memory requirements are also very less. HEAP is resistant to several outsider attacks such as DoS, wormhole, replay, impersonation and man-in-the-middle attacks by making it very difficult for an outsider to propagate any forged packet.

### 3.7. Trust Vector DSR (TVDSR)

Wei Gong et al [19] proposed to use trust vector model based routing protocols. The author proposed an efficacy mechanism which is based on trust evaluation. Each node would evaluate its own trust vector parameters about neighbors through monitoring neighbor's pattern of traffic in network. Then evaluate the performance of the proposed mechanism by modifying Dynamic Source Routing (DSR), so that each node has a dynamic changing trust vector for its neighbor's behaviors. This vector can be normalized into a single trust value that has been provided as evidence for decision making in routing selection process.

Trust vector of node A to B is
$$V(A \rightarrow B) = [_AE_B, {_AK_B}, {_AR_B}]$$
Where $_AE_B, {_AK_B}$ and $_AR_B$ are node A's evaluation of experience, knowledge and recommendation to node B. The normalization of trust vector can be defined as $|V(A \rightarrow B)| = {_AT_B}$.

$(\{_AE_B, {_AK_B}, {_AR_B}\} \in [0,1], \{W_E, W_K, W_R\} \in [0,1])$ where $W_A$ is node a A's trust policy vector, $_AT_B$ is a single trust value of node A on node B corresponding to the normalized trust vector.

$_AE_B$ is node A's evaluation on node B by directly monitoring packet communication of node B. Using (1), $_AE_B$ can be computed by node A :

$$_AE_B = {_AE_B} = \frac{\bar{P}_B}{P_B} = \frac{P_B^{out} - P_B^{B,A}}{P_B^{in} - P_B^{A,B}} \qquad (1)$$

Where $\bar{P}_B$ is the number of packets node B had actually forwarded. $P_B$ is the number of all packets responsible for forwarding. $P_B^{out}$ is all out-coming packets, $P_B^{in}$ is all incoming packets from node B. $P_B^{B,A}$ is packets from source node B to destination node A. $P_B^{A,B}$ packet come from source node A to destination B.

$_AK_B$ is node A's evaluation to node B by directly observing MAC layer link quality between node A and node B on physical layer. Computation formula is as follows:
$$_AK_B = (1-P_{A,B})*(1-P_{B,A})$$
Where, $P_{A,B}$ is packet loss probability from node A to node B, while $P_{B,A}$ is packet loss probability from node B to node A.

$_AR_B$ is node A's evaluation to node B by collecting recommendations about node B from other nodes which should be the neighbor node B. This is given y the equation:
$$_AR_B = \frac{\sum_{C \in \psi} |V(A \rightarrow C)| * |V(C \rightarrow B)|}{\sum_{C \in \psi} |V(A \rightarrow C)|}$$

The time dependent is defined as:
$$_AT_B(t_2) = {_AT_B}(t_1) * e^{-(_AT_B(t_1)\Delta t)^{2k}}$$

$_A T_B(t_1)$ be the trust value of node A to node B at time $t_1$ and $_A T_B(t_2)$ be the decayed value of the same at time $t_2$. $\Delta t = t_2 - t_1$ and k is an integer greater than are equal to 1. Compared to DSR, TVDSR gives higher packet delivery ratio and also reduce the malicious nodes dramatically.

## 4. ANALYSIS AND COMPARISION

Detecting MAC layer misbehavior is a broad area of research that is based on the contention-window MAC protocols.

The contention window cheating is done based on the following metrics.

❖ Correct detection: It is defined as the ratio of the number of misbehaved nodes that are correctly marked by the detection system as suspects to the total number of active misbehaved nodes in the network.
❖ Misdetection: It is defined as the ratio of the number of well behaved nodes that are incorrectly diagnosed as suspects to the total number of well behaved nodes in the network.
❖ Packet delivery ratio: It is the ratio of the data packets successfully delivered to the destination to those generated by the source.
❖ Delay: It is the average delay time of all successfully delivered packets.
❖ Throughput: It is the ratio of the data packets successfully delivered to the destination for each flow to those generated by the source.

Table 1 gives the comparisons of few of the contention window cheating techniques. Misbehaving nodes always choose smaller backoff values for more bandwidth utilization. A combination of contention window misbehaviors along with the adjustment scheme is used to ensure correct misbehavior diagnostics. In DREAM [11] two reaction schemes require minor changes of the existing standard to mitigate the protocol failure or the misuse. The DREAM method is invoked a modified timeout period based on the detection threshold.

Table 1. Comparison of Techniques proposed for detecting selfishness in MANET

| Method | Correct detection | Misdetection | Delivery Ratio | Delay | Throughput |
| --- | --- | --- | --- | --- | --- |
| DREAM | Very good | Good | Very good | Good | Good |
| EDCA | Good | Good | High in TCP | High in UDP | High in TCP |
| KOLMOGOROV-SMIROV(K-S) TEST | Very good | Good | Good | Bad | Good |
| PREDICTABLE RANDOM BACKOFF | Good | Good | Good | Good | Very good |
| EIED BACKOFF ALGORITHM | --- | --- | Good | Good | Very Good |
| DETERMINISTICAL & STASTICAL METHOD | Very Good | Good | Good | Good | Good |
| HEAP | Good | Good | Very good | Bad | Very good |
| TVDSR | Good | Good | Very Good | Bad | Good |

Alberto and Xiaodong [13] developed nonparametric batch and sequential detectors based on the Kolmogorov-Smirov (K-S) statistics that do not require any modification on the existing Sense CSMA/CA protocols. Kolomgorov-smirov can be applied without modification, to a scenario in which the number of competing terminal is changing. Lei and Chadi evaluate the efficiency of the PRB algorithm in mitigating the negative effects of MAC layer selfish misbehavior. DREAM method performs better than any other method. PRB gives higher throughput especially in congested environment. HEAP and TVDSR methods give better delivery ratio compared to any other method. Although the results in the consulted papers always show an improvement of the misbehavior detection, they never considered the correct detection method because they are usually compared with the proposals that do not consider all the metrics.

## 5. CONCLUSION

The recent research efforts have made a lot of progress on contention window cheating misbehavior in MANET. Detecting MAC layer misbehavior is one of the main issues in MANET. In this paper, a comprehensive survey of contention window cheating techniques in MANET that has been presented in the literature. These cheating techniques are modifications of the contention window parameters like backoff time, network allocation vector and SIFS parameters etc. They have a common objective of trying to utilize more bandwidth, reduce the battery power at each node. In many cases it is difficult to compare them directly since each method has a different goal with different assumptions and employs different means to achieve the goal.

## ACKNOWLEDGEMENTS

R. Kalaiarasi is grateful to the Department of Science and Technology (DST-WOS), New Delhi for financial assistance.

International journal of computer science & information Technology (IJCSIT) Vol.2, No.5, October 2010

**R. Kalaiarasi** received her B.Sc degree in Computer Science from Bharathiar University, India in 1996 and Master of Computer Applications from University of Madras in 2004. Currently she is pursuing her Ph.D degree in the Faculty of Science & Humanities, Anna University Chennai, Chennai, India. Her research interests include wireless ad hoc networking, communication networks and distributed resource allocation.

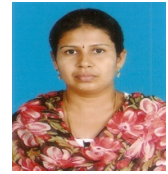

**Getsy S Sara** received her B.E degree with distinction in Electronics & Communication from Bharathiar University, India in 2004 and M.E degree with distinction in Digital Communication Engineering from Anna University, India in 2006. Currently she is pursuing her Ph.D degree in the Faculty of Information & Communication Engineering, Anna University Chennai, India. Her research interests include wireless ad hoc networking, sensor networks, energy efficient routing protocols and communication systems.

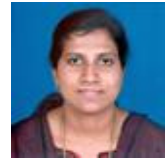

**S Neelavathy Pari** received her B.E degree in Computer Science and Engineering from Kuvempu University, Shimoga in the year 1994 and M.Tech (Honors) Computer Science Engineering from Dr. M.G.R. Educational and Research Institute, India. She is currently working as Lecturer in the Department of Information Technology MIT Campus, Anna University, Chennai, India. Her present research interests include, MANET, sensor network, mobile computing and network security.

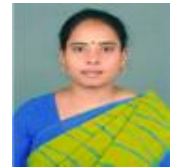

**Dr.D. Sridharan** received his B.Tech. degree and M.E.degree in Electronics Engineering from Madras Institute of Technology, Anna University in the years 1991 and 1993 respectively. He got his Ph.D degree in the Faculty of Information and Communication Engineering, Anna University in 2005. He is currently working as Assistant Professor in the Department of Electronics and Communication Engineering, CEG Campus, Anna University Chennai, Chennai, India. He was awarded the Young Scientist Research Fellowship by SERC of Department of Science and Technology, Government of India. His present research interests include Internet Technology, Network Security, Distributed Computing and Wireless Sensor Networks.

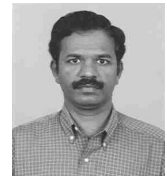